\documentclass[preprint,number,sort&compress]{elsarticle}
\usepackage{graphicx}
\usepackage{subfigure}
\usepackage{amsmath}
\usepackage{amsthm}
\usepackage{amssymb}
\usepackage{verbatim}
\usepackage{multicol}
\usepackage{cite}
\usepackage{accents}

\newlength{\dhatheight}
\newcommand{\doublehat}[1]{%
    \settoheight{\dhatheight}{\ensuremath{\hat{#1}}}%
    \addtolength{\dhatheight}{-0.35ex}%
    \hat{\vphantom{\rule{1pt}{\dhatheight}}%
    \smash{\hat{#1}}}}

\def\eg{{\it e.g.}}
\def\ie{{\it i.e.}}

\begin{document}
\title{Calculating a confidence interval on the sum of binned leakage}

\author[su,rit]{I.~Ruchlin\fnref{fn1,fn2,fn3}\corref{cor1}}
\ead{ixr5289@rit.edu}
\cortext[cor1]{Corresponding author}
\fntext[fn1]{Current address is at Rochester Institute of Technology.}
\fntext[fn2]{Phone: +1 585 475 2498}
\fntext[fn3]{Fax: +1 585 475 7340}

\author[su]{R.W.~Schnee}
\ead{rwschnee@phy.syr.edu}

\address[su]{Department of Physics, Syracuse University, Syracuse, New York 13244, USA}
\address[rit]{Center for Computational Relativity and Gravitation, Rochester Institute of Technology, 85 Lomb Memorial Drive, Rochester, New York 14623, USA}

\begin{abstract}
Calculating the expected number of misclassified outcomes is a standard problem of particular interest for rare-event searches.
The Clopper-Pearson method allows calculation of classical confidence intervals on the amount of misclassification if data are all drawn from the same binomial probability distribution.  However,
 data is often better described by breaking it up into several bins, each represented by a different binomial distribution.  
  We describe and provide an algorithm for calculating a classical confidence interval on the expected total number of misclassified events from several bins, based on calibration data with the same probability of misclassification on a bin-by-bin basis.
 Our method avoids a computationally intensive multidimensional search 
   by introducing a Lagrange multiplier and performing standard root finding.  
   This method has only quadratic time complexity as the number of bins, and produces confidence intervals that are only slightly conservative.
\end{abstract}

\begin{keyword}
confidence interval \sep likelihood ratio \sep binomial distribution
\end{keyword}

\maketitle

\section{Introduction}

Many real-world processes can assume one of two possible outcomes; each independent 
trial or observation can 
be classified as either ``success'' or ``failure,'' with the probability of success $p$.  All such trials
can be bundled together 
to form a single experiment with $x$ successes out of a total of $n$ trials.  If the experiments are repeated many times, the relative frequency of successes in each experiment follows the binomial distribution (\eg~\citep{StatisticalDataAnalysis}).

For a measurement of $x$ and $n$, 
the best estimate $P = x/n$ of the true success probability $p$ can be calculated.
Since the ratio $P/(1 - P)$ is the best estimate of the expected ratio of successes to failures, the best estimate of the number of successes $Y$ of a second experiment that has the same success probability and a known number of failures $b$ is 
\begin{equation}
Y =  \frac{P}{1 - P} b \; .
  \label{eqn:expected_leakage}
\end{equation}
Furthermore, methods such as Clopper-Pearson's~\citep{ConfidenceInterval}
provide a (frequentist, or classical)
confidence interval $[P_{\mathrm{low}}, P_{\mathrm{high}}]$ with probability content $\beta$ 
such that the fraction of experiments with $P_{\mathrm{low}}\leq p \leq  P_{\mathrm{high}}$ is $\approx\beta$ (with the lack of exact equality due to the discrete nature of binomial distribution; see \eg~\citep{IntervalEstimation} for comparisons of various methods).
By extension, these methods  
may also be used to calculate the confidence interval 
$[Y_{\mathrm{low}}, Y_{\mathrm{high}}]$ on the expected number of successes of the second experiment.

Such estimates may be particularly useful for characterizing backgrounds for rare-event searches.
A given background event may have some probability $p$ to be misclassified as a signal event.  A first, ``calibration'' experiment may allow estimation of  $p$ based on the number of events $n$ and the number $x$ misclassified as signal (the ``leakage'').
A second, ``search'' experiment may provide a measurement of the number of correctly identified background events $b$.  If background events in both experiments have the same probability of correct classification, the expected number of misclassified events $Y$ and a confidence interval $[Y_{\mathrm{low}}, Y_{\mathrm{high}}]$ on the expected number may be determined.

Often, however, in order for events in the calibration and search both to have the same probability of misclassification $p$, events with different characteristics (\eg\ energy, position, detector, or pixel) must be considered separately, resulting in $m$ separate bins of events for both calibration and search.
In the $i^{th}$ calibration bin there are $x_i$ misclassified events out of the total $n_i$ calibration events, resulting in a best estimate $P_i = x_i/n_i$ of the misclassification probability for events in that bin.  
For the search data, the number of correctly classified events in the $i^{th}$ bin, $b_i$, is known.  
If the true misclassification probability, $p_i$, of an event in the $i^{th}$ bin is the same for both calibration and search data,
the best estimate for the total expected number of misclassified events $Y$ is 

\begin{equation}
 Y = \sum_i^m \frac{P_i}{1 - P_i} b_i  \equiv f(\mathbf{P}) \; ,
  \label{eqn:total_expected_leakage}
\end{equation}
where $\mathbf{P} \equiv \{ {P_i} \}$.  

The likelihood ${\mathcal{L}}$ that $x_i$ events out of the total $n_i$ calibration events in each bin are misclassified is simply the product of the binomial probabilities, with
\begin{equation}
  \mathcal{L} \propto \prod_i^m P_{i}^{x_i} (1-P_{i})^{n_i-x_i} \; .
  \label{eqn:likelihood_function}
\end{equation}
The global maximum $\hat{\mathcal{L}}$ of the likelihood is trivially given by the set $ \hat{\mathbf{P}}\equiv \{ \hat{P_i} \} = \{ x_i / n_i \}$ for all $i$.  
Note that it is possible to estimate the expected leakage only if no calibration bin has zero total events (i.e.\ $n_i \neq 0$ for all $i$).
  Substitution of  the set $\{ \hat{P_i} \}$ into Equation~\ref{eqn:total_expected_leakage}
yields the most likely value of the total expected leakage $\hat{Y}=f(\hat{\mathbf{P}}) $.

Unfortunately, for the case with multiple bins,
 most existing methods
 cannot be used to calculate a confidence interval on the total expected leakage.
Here we describe a method and provide a practical algorithm for this problem.
We use the ``Unified Approach'' described by Feldman and Cousins~\citep{fc} (see also \eg~\citep{kendall})
extended to deal with nuisance variables by means of the profile likelihood~\citep{FeldmanProfile} without the large-sample approximation used in \eg~\citep{StatisticalDataAnalysis,Rolke:2004mj}; here  $\mathbf{P}$ are nuisance variables
since they are unknown variables for which we are not setting a confidence interval.

\section{Method}

For every considered value of $Y_0$, we calculate the profile likelihood
\begin{equation}
  \Lambda \equiv \frac{\mathcal{L}\left ( Y_0 | \mathbf{n}, \mathbf{x}, \mathbf{b}, \doublehat{\mathbf{P}} \right )}{\hat{\mathcal{L}}\left ( \hat{Y} | \mathbf{n}, \mathbf{x}, \mathbf{b}, \hat{\mathbf{P}} \right )} \; ,
  \label{eqn:likelihood_ratio}
\end{equation}
where  $\mathbf{n} = \{n_i\}$, $\mathbf{x} = \{x_i\}$, and $\mathbf{b} = \{b_i\}$ are the data, and $\doublehat{\mathbf{P}}$ is  the combination of $P_i$, found by a search described in Sect.~\ref{formulation}, that maximizes the likelihood for the value of $Y_0$ under test.
Asymptotically (and far from physical boundaries), the distribution of $-2 \ln(\Lambda)$ is $\chi^2$-distributed with 1 degree of freedom~\citep{WilksTheorem}, but
more accurate results may be obtained by determining the expected distribution 
by Monte Carlo simulation.  For each simulated experiment, $\mathbf{x}$ is randomly determined based on the $\doublehat{\mathbf{P}}$ found above.  
For each, the best-fit values $\hat{\mathbf{P}}_{\text{MC}}$ and $\doublehat{\mathbf{P}}_{\text{MC}}$ are found, and then the ratio 

\begin{equation*}
  \Lambda_{\text{MC}} \equiv \frac{\mathcal{L}\left( Y_0 | \mathbf{n}, \mathbf{x}, \mathbf{b}, \doublehat{\mathbf{P}}_{\text{MC}}\right)}{\hat{\mathcal{L}}\left(\hat{Y} | \mathbf{n}, \mathbf{x}, \mathbf{b}, \hat{\mathbf{P}}_{\text{MC}}\right)}
\end{equation*}
is calculated.  
If $\Lambda$ is larger than $1-\beta$ of the simulated $\Lambda_\text{MC}$ ratios, then $Y_0$ is included in the confidence interval of probability content $\beta$.  Since the distributions follow the binomial distribution, the uncertainties $\propto 1/\sqrt{N_\text{MC}}$, the inverse of the square root of the number of experiments.  Thus, to achieve a relative tolerance $t$, conduct $N_{\text{MC}} = t^{-2}$ Monte Carlo simulations.  A root-finding algorithm hunts for the smallest and largest values of $Y_0$ that are allowed in order to return the desired confidence interval $[Y_{\mathrm{low}}, Y_{\mathrm{high}}]$.

\subsection{Formulation}
\label{formulation}

 A multiparameter function minimizer, such as MINUIT~\citep{minuit}, could be implemented to hunt for the 
 combination of probabilities $P_{i}$ that maximize Equation~\ref{eqn:likelihood_function} subject to the constraint of Equation~\ref{eqn:total_expected_leakage}.
However,  this method may have 
exponential time complexity in the worst case~\citep{SimplexComplexity}, making it unfeasible for the analysis of more than a few bins.  Furthermore, there would be some risk of missing the global maximum.
Instead, the combination of binomial probabilities that maximize Equation~\ref{eqn:likelihood_function} subject to the constraint of Equation~\ref{eqn:total_expected_leakage} can be found by introducing a Lagrange multiplier, $\lambda$, and solving

\begin{equation*}
  \frac{\partial}{\partial P_i} \left[ \ln( \mathcal{L}(P_i)) + \lambda (f(\mathbf{P}) - Y_0) \right] = 0 \; ,
\end{equation*}
where $Y_0$ is a given constant.  A little algebra yields the solution to this equation for each bin $i$:

\begin{equation}
  P_i = \frac{n_i + x_i - \lambda b_i \pm \sqrt{(\lambda b_i - n_i - x_i)^2 - 4 n_i x_i}}{2 n_i} \; ,
  \label{eqn:probability_lambda}
\end{equation}
while substituting back into Equation~\ref{eqn:total_expected_leakage} yields an equation for the Lagrange multiplier:

\begin{equation}
  Y_0 = \sum_i^m b_i \frac{n_i + x_i - \lambda b_i \pm \sqrt{(\lambda b_i - n_i - x_i)^2 - 4 n_i x_i}}{n_i - x_i + \lambda b_i \mp \sqrt{(\lambda b_i - n_i - x_i)^2 - 4 n_i x_i}} 
  \equiv \sum_i^m b_i Y_{0i}\; .
  \label{eqn:total_expected_leakage_lambda}
\end{equation}

Equation~\ref{eqn:total_expected_leakage_lambda} is really $2^m$ separate equations, depending on the the signs of \emph{each} $\pm$ term. 
One of the $2^m$  solutions yields the value of $\lambda$ that gives the most likely combination of binomial probabilities (i.e.\ $\doublehat{\mathbf{P}}$) for the desired total expected leakage $Y_0$. 
Fortunately, further analysis reveals a significant reduction in the number of viable solutions.

For any bin with $b_i \neq 0$,
\begin{equation*}
 \lambda > c_i \equiv  \frac{n_i + x_i - 2 \sqrt{n_i x_i}}{b_i} 
\end{equation*}
is unphysical, producing imaginary or negative probabilities.
Since $\lambda$ must be physical for all bins, $\lambda$ is required to be less than or equal to the smallest $c_i$, \ie\ $\lambda \leq \inf\{c_i\} \equiv \lambda_c$.  For any bin with $b_i = 0$, $c_i \to \infty$ so that it places no constraint on $\lambda$.

Table~\ref{tab:probability_roots} lists the different limiting values of $\lambda$ and their corresponding values of $P_i$ from Equation~\ref{eqn:probability_lambda}.  The lower limit on the confidence interval must have $P_i \leq x_i / n_i$ for each bin.  Therefore, Table~\ref{tab:probability_roots} indicates that the 
solution 
must use the negative root for each bin, reducing the problem from searching among $2^m$ solutions
to solving a single equation.

\begin{table}[t]
  \caption{Summarized analysis of the behavior of Equation~\ref{eqn:probability_lambda}.} 
  \begin{center}
  \begin{tabular}{r @{ $\lambda$ } l r @{ $P_i$ } l  r @{ $P_i$ } l }
    \hline
    \hline
    \multicolumn{2}{c}{} & \multicolumn{2}{c}{Positive Root}  & \multicolumn{2}{c}{Negative Root} \\
    \hline
    & $< 0$ &                 & $ > 1$                                     & $0 < $ & $ < x_i / n_i$ \\
    & $= 0$ &                 & $ = 1$                                     &              & $ = x_i / n_i$ \\
       $0 < $ & $ < c_i$ & $\sqrt{x_i/n_i} < $ & $ < 1$ & $x_i/n_i < $ & $ < \sqrt{x_i/n_i}$ \\
    \hline
  \end{tabular}
  \end{center}
  \label{tab:probability_roots}
\end{table}

It is easiest to understand the viable solutions for the confidence interval's upper bound by first noting that, other than the constraint of Equation~\ref{eqn:total_expected_leakage}, each term in 
\begin{equation*}
\ln (\mathcal{L}) = \sum_i^m \ln  (\mathcal{L}_i)
\end{equation*}
is independent.  
For each term, $\ln  (\mathcal{L}_i)$ decreases monotonically with increasing $P_i>\hat{P}_i$, 
and there is an inflection point at $P_i=P_i(\lambda=c_i)$, with $\ln  (\mathcal{L}_i)$ decreasing ever more slowly for larger $P_i$.

For any bins $i$ and $j$, it can be shown that
\begin{equation*}
\left.  \frac{\partial \ln(\mathcal{L}) }
  {\partial 
  Y_{0i}} \right|_{P_i=\doublehat{P}_i }= \left. \frac{\partial \ln(\mathcal{L}) }
  {\partial
  Y_{0j}} \right|_{P_j=\doublehat{P}_j }\; .
\end{equation*}
This relation is to be expected.  If, instead, the left term were larger (smaller) than the right term,
a more likely combination with the same total $Y_0$ could be found by decreasing $Y_{0i}$ ($Y_{0j}$) and increasing the other by the same amount.

\begin{figure}[t]
  \centering
  \includegraphics[scale=0.35]{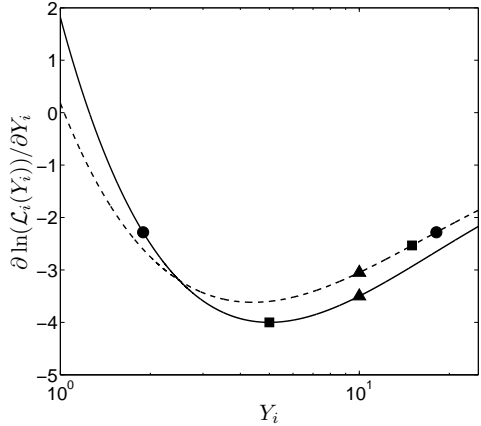}
  \caption{Visualization to demonstrate that the total likelihood is not maximized if more than one bin uses the positive root in Equation~\ref{eqn:total_expected_leakage}.  Any points to the right of each curve's minimum are given by the positive root, while points at or to the left of the minimum are given by the negative root.  Suppose the desired total expected leakage is $Y_0 = 20$, with one possible combination, $Y_{01} = Y_{02} = 10$,  shown as the triangles.  
Since $\partial \ln(\mathcal{L}) / \partial Y_{0i}$ must increase with increasing $Y_{0i}$ for both curves to the right of their minima, 
a more likely combination also resulting in $Y_0 = 20$ can be found by 
  decreasing the expected leakage in the bin with the more negative value of $\partial \ln(\mathcal{L}) / \partial Y_{0i}$ to the minimum while increasing the expected leakage in the other (squares).
The maximum of the likelihood may be found by continuing to decrease the expected leakage in the bin with the more negative value of $\partial \ln(\mathcal{L}) / \partial Y_{0i}$ until both bins have equal values of  $\partial \ln(\mathcal{L}) / \partial Y_{0i}$ (circles).
  }
  \label{fig:root_proof}
\end{figure}

In a similar way, it may be shown that the combination of probabilities $\doublehat{\mathbf{P}}$ that maximize the likelihood for a given total expected leakage $Y_0$ never includes more than one bin with $P_i>P_i(\lambda=c_i)$, and hence more than one bin using the positive root of Equation~\ref{eqn:total_expected_leakage}.  Figure~\ref{fig:root_proof} helps visualize the reasoning.
Consider any two bins $i$ and $j$ that both use the positive roots and hence have $P_i>P_i(\lambda=c_i)$, $P_j>P_j(\lambda=c_j)$.
If $\partial \ln(\mathcal{L}) / \partial Y_{0i} |_{Y_{0i} = k} \geq  \partial \ln(\mathcal{L}) / \partial Y_{0j} |_{Y_{0j} = k}$ then $\partial \ln(\mathcal{L}) / \partial Y_{0i} |_{Y_{0i} = k+\delta} \geq  \partial \ln(\mathcal{L}) / \partial Y_{0j} |_{Y_{0j} = k-\delta}$ for $\delta = k - Y_{0j}(c_j)$, since $\partial \ln(\mathcal{L}) / \partial Y_{0i} $ becomes larger as $Y_{0i}$ is increased, and $\partial \ln(\mathcal{L}) / \partial Y_{0j}$ becomes smaller as $Y_{0j}$ is decreased towards the inflection point. 
As a result, for a given total leakage, the likelihood is never maximized by having two bins use the positive roots of  Equation~\ref{eqn:total_expected_leakage}.
Since a multi-bin system can be examined in $2$-bin pieces, it follows that it is necessary to solve Equation~\ref{eqn:total_expected_leakage_lambda} only for $m+1$ cases: the case 
with all negative roots, and also the $m$ cases
with one positive root and 
$m-1$ negative roots.  Thus, the total number of equations to be solved is reduced from $2^m$ to $m+1$.

A standard root-finding algorithm may be used to hunt for a physical solution to each of these $m+1$ equations.  Our implementation uses the Van Wijngaarden-Dekker-Brent Method~\citep{NumericalRecipes} which combines the secant method, bisection method, and inverse quadratic interpolation to find bracketed roots of a given function---in this case, appropriate 
$\lambda$ for a given $Y_0$.

\subsection{Benchmarking}

All tests were conducted with an Intel Core i5 M430 processor and 6 GB of RAM.  Since each additional bin increases the length of all of the data storage vectors by one element, the RAM usage increases linearly with the number of bins.
 The estimated RAM usage is $16.69 \pm 0.01$ kB per bin.  
 For these tests, five measurements of each quantity were taken and their averages were plotted.  The error bars represent the standard deviations of those means.

\begin{figure}[th]
  \centering
  \includegraphics[scale=0.5, angle=-90]{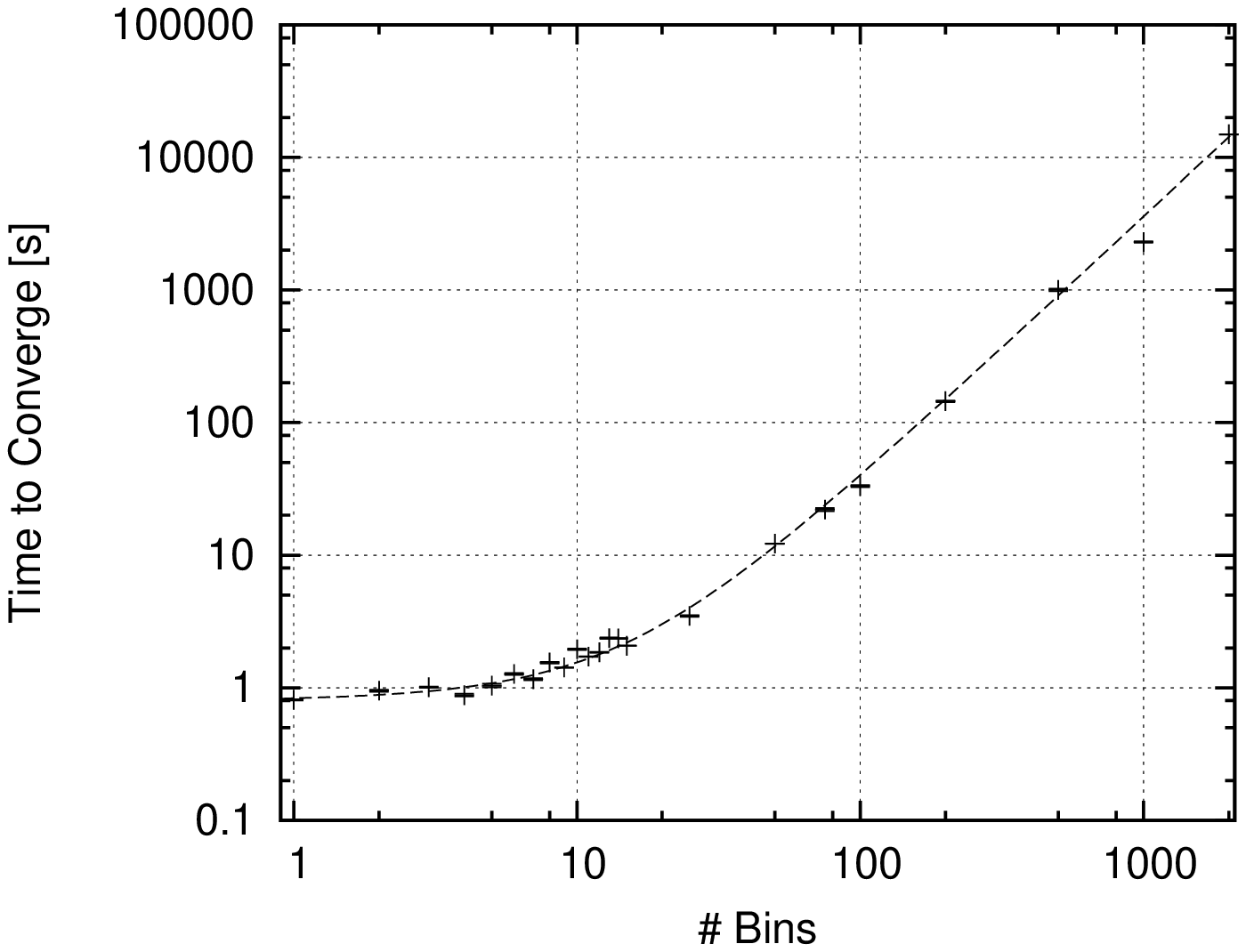}
  \includegraphics[scale=0.5, angle=-90]{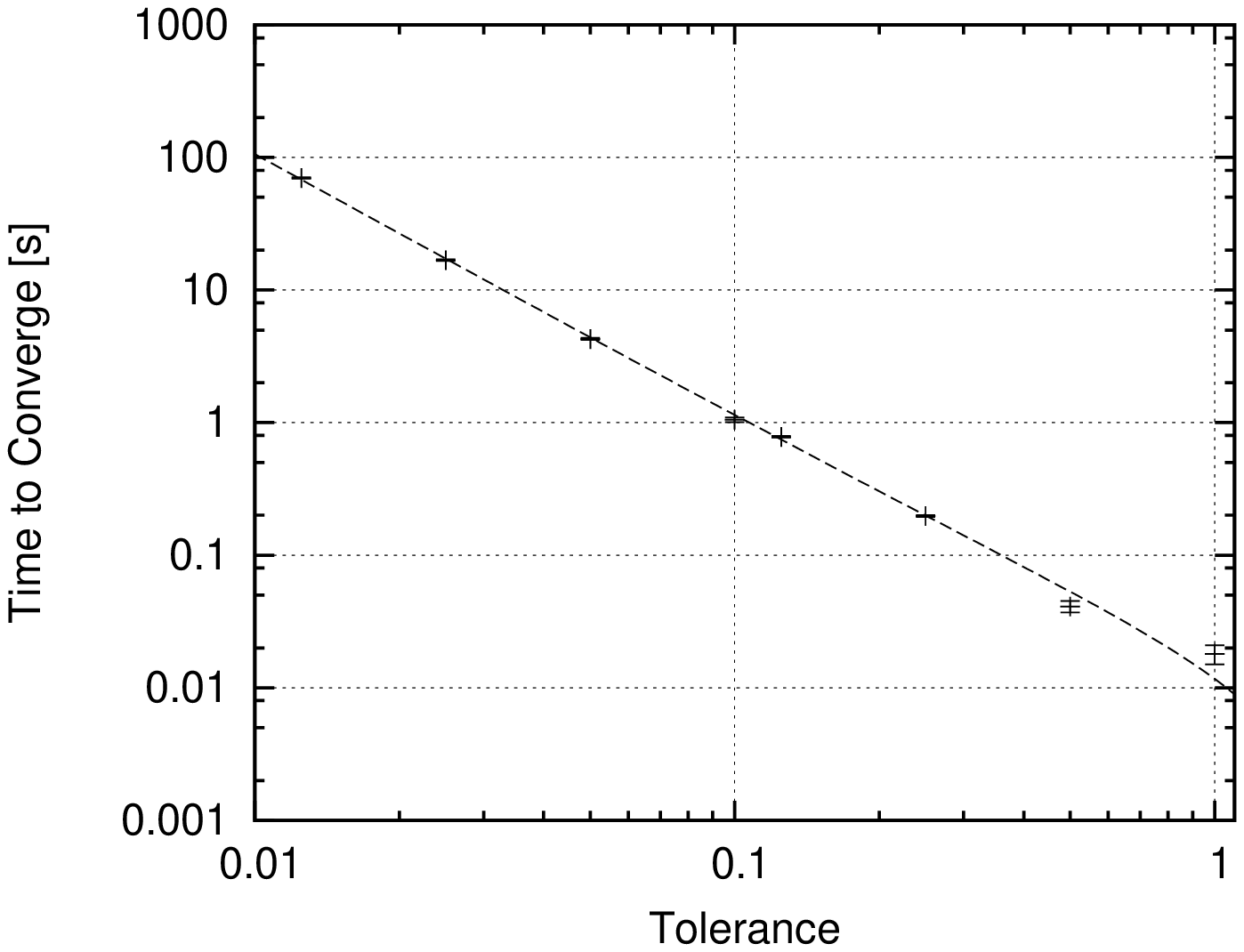}
  \caption{{\it Top}: Run time as a function of the number of bins $m$, each with $n_i=100$ calibration events, $x_i=5$ leaking calibration events, and $b_i=10$ correctly tagged events from the search data, for tolerance $t = 0.1$.  The data fit time complexity $\mathcal{O}(m^2)$.
{\it Bottom}: Run time as a function of tolerance $t$, for $m=5$ bins of data identical to that used in the \textit{top} plot.  The data fit time complexity $2^{-\mathcal{O}(\log(t))}$, polynomial time. }
  \label{fig:time_vs_bin}
  \label{fig:time_vs_tol}
\end{figure}

For $m$ bins, 
Equation~\ref{eqn:total_expected_leakage} consists of $m$ terms that must be calculated for a given value of $\lambda$.
To find both the upper and lower limits, $m+1$ equations each with $m$ terms must be solved.
Therefore,
finding the limits of the confidence interval as a function of the number of bins 
has
time complexity $\mathcal{O}(m^2)$, 
as shown in Figure~\ref{fig:time_vs_bin}.
The number of Monte Carlo simulations is inversely proportional to the square of the desired relative tolerance.  Therefore, as tolerance decreases, the number of basic operations increases polynomially,
as shown in Figure~\ref{fig:time_vs_tol}.

\subsection{Coverage Tests}
\label{Coverage_Tests}

By construction, the confidence interval includes the true expected leakage a fraction $\beta$ of the time if the true leakage probabilities $\mathbf{p} = \doublehat{\mathbf{P}}$, \ie\ the method provides correct ``coverage'' for the most likely combination of probabilities for each total expected search leakage $Y_0$.  
For other values of the true leakage probabilities, the coverage is not exact.  
However, since $\doublehat{\mathbf{P}}$ is the combination of nuisance parameters that maximizes the likelihood for the data, it may be expected that the coverage is nearly correct for other combinations of true leakage probabilities, and that there is usually slight over-coverage (\eg\ a claimed 90\% confidence interval may contain the true parameters slightly more than 90\% of the time for some combinations of parameters). 

\begin{figure}[t]
  \centering
  \includegraphics[scale=0.5, angle=-90]{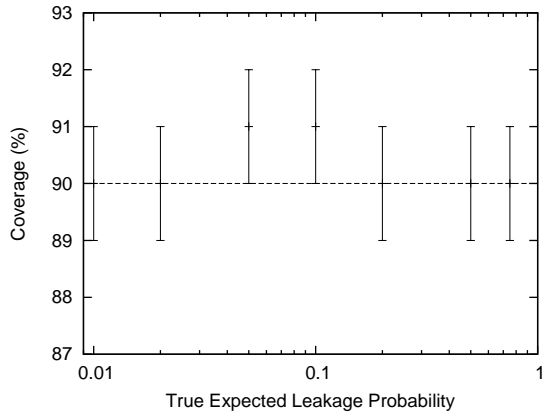}
  \caption{The percentage of simulated 90\% confidence intervals 
  that include the true expected
  leakage given the probability that an event will leak.  
  Each test uses 
  3 bins, with $n_i = 1000$, $b_i = 10$, and $x_i$ thrown from a binomial distribution given $n_i$ and the true expected leakage.
  }
  \label{fig:coverage_vs_tel}
\end{figure}

To test this algorithm for coverage, several simulated experiments were performed.  For each set of tests, values of $\mathbf{n}$, $\mathbf{b}$, and $\mathbf{p}$ were chosen as listed in Table~\ref{tab:coverage_tests}
or in the caption to Figure~\ref{fig:coverage_vs_tel}.
The calibration leakage $\mathbf{x}$ was simulated
from the binomial distribution, given $\mathbf{n}$ and $\mathbf{p}$,
and the program found the most likely total expected leakage and corresponding 90\% confidence interval.  For each set of tests, this simulation was repeated $10000$ times in order to  
determine what fraction of the intervals contained the true expected leakage. 
The results, shown in Figure~\ref{fig:coverage_vs_tel} and Table~\ref{tab:coverage_tests},
  suggest that the algorithm produces confidence intervals with approximate coverage and slight conservatism, as expected.

\begin{table}[ht]
  \caption{Total number of calibration events $\mathbf{n}$, expected calibration leakage $\langle\mathbf{x}\rangle$, number of correctly tagged search events $\mathbf{b}$, and percentage of the time that the true expected search leakage was in the resulting 90\% confidence interval, for the various cases described.}
  \begin{center}
  \begin{tabular}{ c c c c l}
    \hline
    \hline
    $\mathbf{n}$ & $\langle\mathbf{x}\rangle$ & $\mathbf{b}$ & \% In
    & Description \\
    \hline
    $(10^3, 10^3)$ & $(1, 100)$ & $(1, 100)$ & $93 \pm 1$ &
    bin with large $\langle x_i \rangle$,
     $b_i$ \\
    $(10^3,10^3,10^3)$ & $(500,5,5)$ & $(10,10,10)$ & $90 \pm 1$ &
    bin with
    large $\langle x_i \rangle$  \\
    $(10, 10^3, 10^3)$ & $(5, 5, 5)$ & $(10^4, 10, 10)$ & $89 \pm 1$ &
    bin with $n_i \ll b_i$ \\
    $(10^5, 10^3, 10^3)$ & $(5, 5, 5)$ & $(10, 10, 10)$ & $90 \pm 1$ &
    bin with $n_i \gg b_i$ \\
    $(10^3, 10^3, 10^3)$ & $(100, 50, 30)$ & $(10, 10, 10)$ & $91 \pm 1$ & wide range of $\langle \mathbf{x} \rangle$ \\
    \hline
  \end{tabular}
  \end{center}
  \label{tab:coverage_tests}
\end{table}

\section{Application to Dark Matter Searches}

Many collaborations around the world are attempting to make direct detections of dark matter particles (see \eg~\citep{spooner2007review,bertonePDM,SchneeTASI2009}). Dark matter is hypothesized to constitute the majority of the universe's mass, in an effort to explain many astrophysical observations~\citep{WMAP}.  Since dark matter particles interact very weakly with normal matter, dark matter detectors must be very sensitive, which makes the rejection of background interactions an important task.

\begin{figure}[hb]
  \centering
    \includegraphics[scale=0.27]{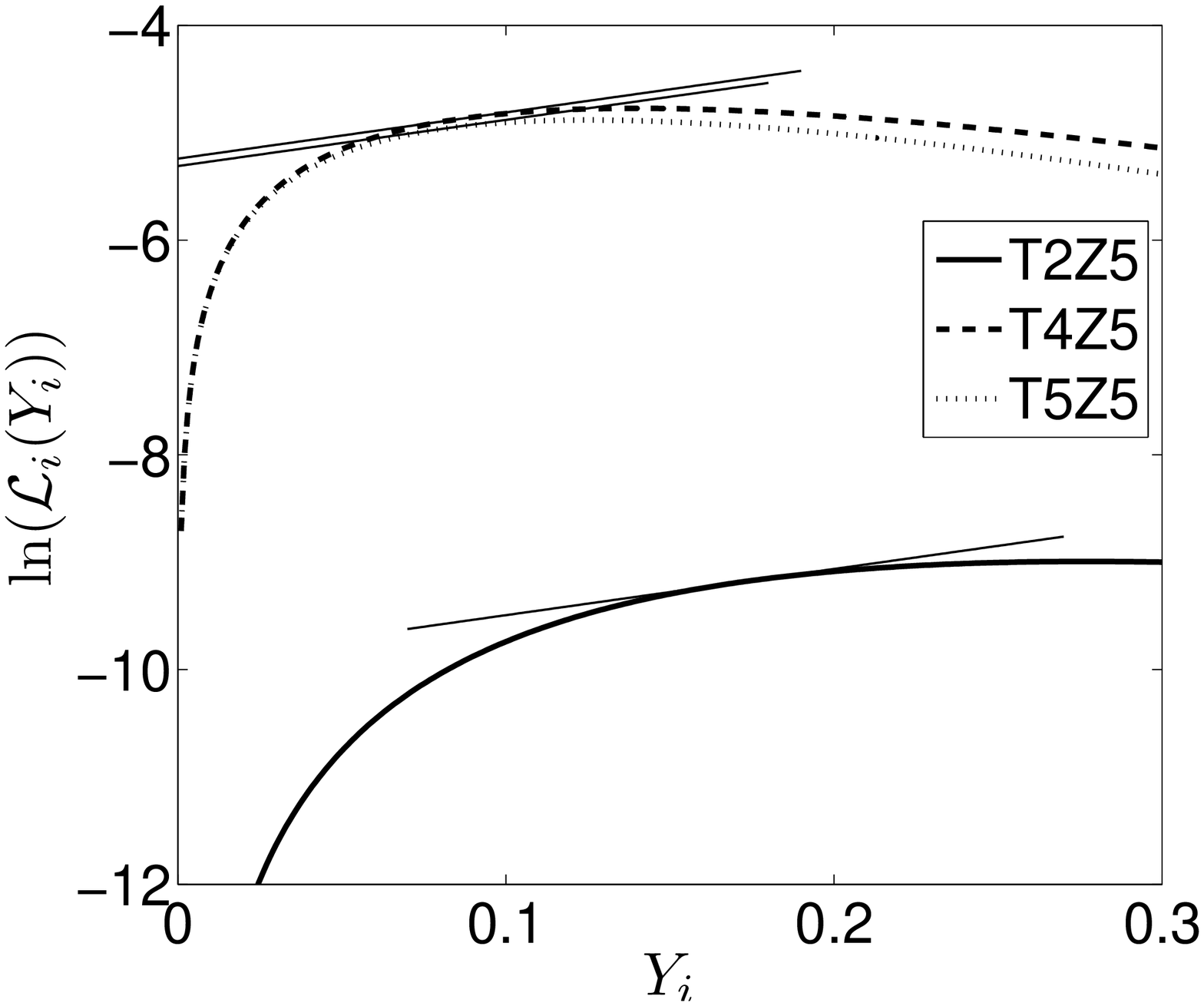}
  \includegraphics[scale=0.27]{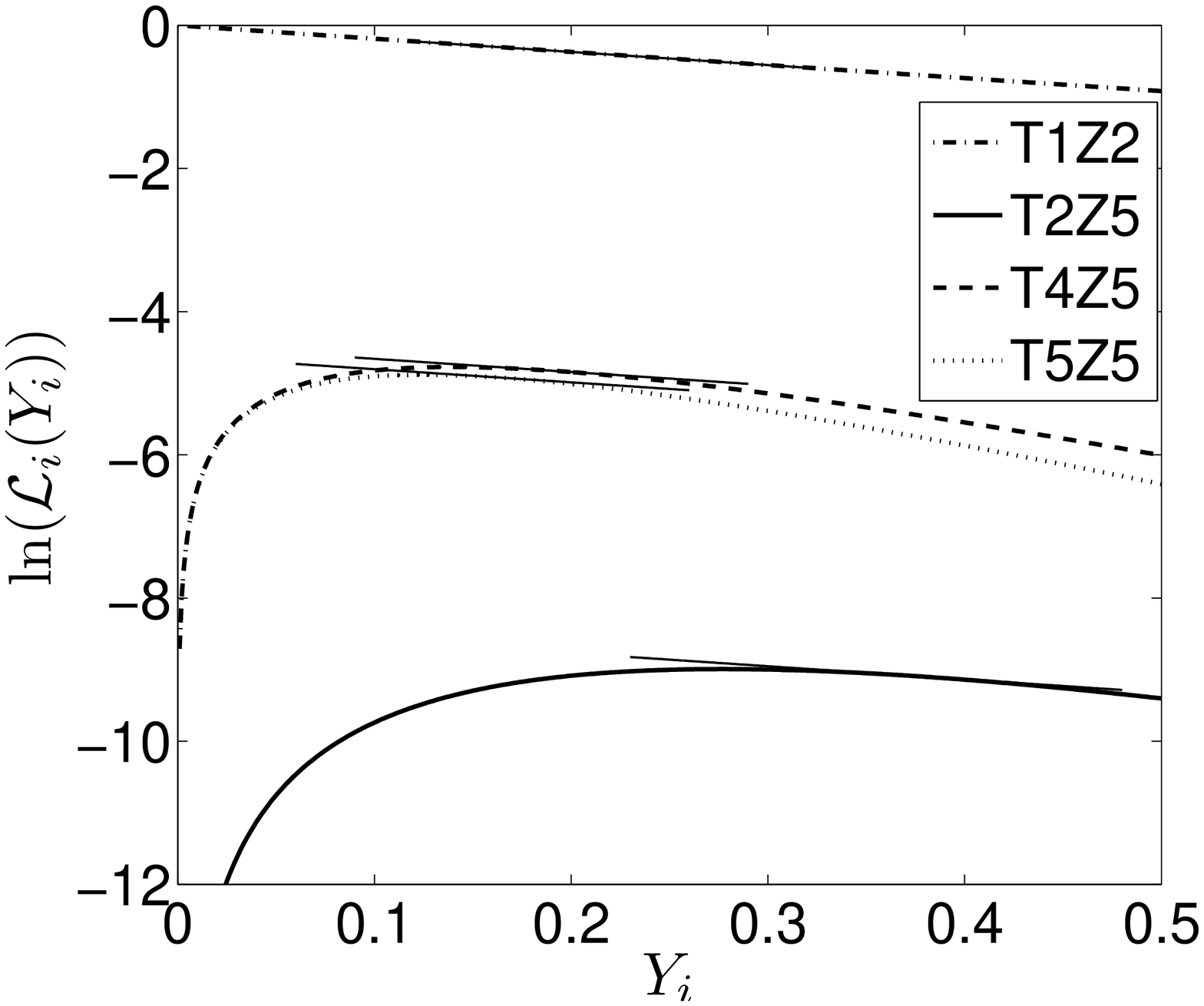}
  \caption{Plot of the likelihood ${\mathcal{L}_i}$ of the expected leakage $Y_{0i}$ in each bin versus the expected leakage in each bin for the CDMS Run $125$ c58.  Of the twelve bins under consideration, only the three with $x_i>0$ (T2Z5: solid, T4Z5: dashes, and T5Z5: dots) have non-zero expected leakage solutions for the lower limit and only 
  one additional bin (T1Z2: dash-dots) has $Y_{0i}>0$ for the upper limit.  The straight lines are tangent to each bin at the points corresponding to the most likely combination of expected leakages, yielding the 68\% lower (\textit{left}) and upper (\textit{right}) limits of the confidence interval.  
 For the upper limit, the overall likelihood of the configuration is decreased the least by increasing the expected leakage in the detector with the worst statistics---T1Z2.  As expected, the tangent lines are all parallel.}
  \label{fig:ln_Yi}
\end{figure}

The CDMS II collaboration recently published results in which two candidate dark matter events appeared in their signal region.  However, with an expected background of $0.9 \pm 0.2$ events, it was 
not statistically significant evidence of dark matter particle detection~\citep{CDMSscience}.  
The background estimate was calculated with a Bayesian technique~\citep{Jeff_thesis}
based on three independent ``calibration'' data sets, two of which are too complicated to be considered here.  The third ``calibration'' set  (see Table~\ref{tab:c58_data}) provided a background estimate only for those detectors not on the top or bottom of a detector stack; 
this background estimate totaled $0.65^{+0.46}_{-0.29}$ events (stat.) $\pm 0.13$ events systematic uncertainty, with a slight expected overcoverage based on simulations.  The same data, when analyzed by the method described above, give a total expected leakage of $0.54^{+0.41}_{-0.20}$ events, confirming the expected slight overcoverage in  the CDMS II estimate.  A visual representation of the upper and lower bound solutions are shown in Figure~\ref{fig:ln_Yi}.  Because the other calibration sets result in smaller uncertainties than the third calibration set, including our revised estimate of the leakage has a negligible effect on the total estimated background.

\begin{table}[ht]
  \caption{Data from the final run of the CDMS II experiment.  Each detector is taken to be a single bin.  Two bins (the end-cap detectors T3Z6 and T4Z6) have been excluded.}
  \begin{center}
  \begin{tabular}{ c c c c }
    \hline
    \hline
    Detector Designation & $n_i$ & $x_i$ & $b_i$ \\
    \hline
    T1Z2 & 28 & 0 & 15 \\
    T1Z5 & 49 & 0 & 8 \\
    T2Z3 & 45 & 0 & 8 \\
    T2Z5 & 67 & 2 & 9 \\
    T3Z2 & 50 & 0 & 2 \\
    T3Z4 & 48 & 0 & 7 \\
    T3Z5 & 29 & 0 & 4 \\
    T4Z2 & 41 & 0 & 5 \\
    T4Z4 & 31 & 0 & 6 \\
    T4Z5 & 44 & 1 & 6 \\
    T5Z4 & 59 & 0 & 6 \\
    T5Z5 & 49 & 1 & 6 \\
    \hline
  \end{tabular}
  \end{center}
  \label{tab:c58_data}
\end{table}

\section{Discussion}

The method described here 
solves the problem of calculating confidence intervals on leakage for several bins.  It has quadratic time complexity, so that even problems with a tremendous number of bins may be handled.  Furthermore, it produces intervals that are only slightly conservative, but exhibit full coverage.

A more rigorous and detailed discussion of the topics presented in this paper has been compiled into a technical note, along with a fully coded version of the algorithm~\citep{IntervalWebsite}.  

\section{Acknowledgements}

This work is supported in part by the 
National Science Foundation (Grant No.\ PHY-0855525).
We thank Jeff Filippini for his comments.

\bibliographystyle{unsrt}
\bibliography{bibliography}

\end{document}